\documentclass[aps,prd,twocolumn,superscriptaddress,showpacs,showkeys]{revtex4}
\usepackage{epsfig}
\usepackage{graphicx}
\usepackage{amsmath,amssymb}

\newcommand{\bastar}{\begin{eqnarray*}}
\newcommand{\eastar}{\end{eqnarray*}}
\newskip\humongous \humongous=0pt plus 1000pt minus 1000pt

\newif\ifdtup

\relax
\newcommand{\bea}{\begin{eqnarray}}
\newcommand{\eea}{\end{eqnarray}}
\newcommand{\X}{\vec X}
\newcommand{\W}{\vec W}
\newcommand{\pro}{\partial}
\newcommand{\pd}{\partial}
\newcommand{\n}{\hat n}

\newcommand{\mn}{\mu\nu}

\newcommand{\F}{\vec F}

\newcommand{\hF}{\hat F}

\newcommand{\A}{\vec A}
\newcommand{\hA}{\hat A}
\newcommand{\cA}{\cal A}
\newcommand{\cC}{\cal C}
\newcommand{\tA}{\tilde A}
\newcommand{\tC}{\tilde C}

\newcommand{\vsig}{\vec \sigma}
\newcommand{\vlam}{\vec \lambda}
\newcommand{\hD}{\hat D}

\newcommand{\nn}{\nonumber}

\newcommand{\bPsi}{\bar \Psi}
\newcommand{\bp}{\bar p}
\begin{document}
\title{Probing Two types of Gluon Jets at LHC}
\bigskip
\author{Y. M. Cho}
\email{ymcho0416@gmail.com}
\affiliation{Institute of Modern Physics, Chinese Academy of
Science, Lanzhou 730000, China}
\affiliation{School of Physics and Astronomy,
Seoul National University, Seoul 08826, Korea}
\affiliation{Center for Quantum Spacetime, Sogang 
University, Seoul 04107, Korea}  
\author{Franklin H. Cho}
\affiliation{Institute for Quantum Computing,
Department of Physics and Astronomy  \\
University of Waterloo, Ontario N2L 3G1, Canada}
\author{Pengming Zhang}
\email{zhpm@impcas.ac.cn}
\affiliation{Institute of Modern Physics, Chinese Academy of
Science, Lanzhou 730000, China}

\begin{abstract}
We propose a simple way to test the Abelian decomposition 
of QCD, the existence of two types of gluons, experimentally 
at LHC. The Abelian decomposition decomposes the gluons 
to the color neutral neurons and colored  chromons gauge independently. This refines the Feynman diagram in a way 
that the color conservation is explicit, and generalizes 
the quark model to the quark and chromon model. We 
predict that the neuron jet has the color factor 3/4 and 
has a sharpest jet radius and smallest particle multiplicity, 
while the chromon jet with the color factor 9/4 remains 
the broadest jet. Moreover, the neuron jet has a distinct 
color flow which forms an ideal color dipole, while the quark 
and chromon jets have distorted dipole pattern.    
\end{abstract}
\pacs{12.38.-t, 12.38.Aw, 11.15.-q, 11.15.Tk}
\keywords{Abelian decomposition, two types of gluons, 
neuron, chromon, decomposition of Feynman diagram 
in QCD, neuron jet, chromon jet, color factors of neuron 
and chromon jets, quark and chromon model}
\maketitle

A common misunderstanding on QCD is that 
the non-Abelian gauge symmetry is so tight that 
it defines the theory almost uniquely, and thus 
does not allow any simplification. The Abelian 
decomposition of QCD tells that this popular wisdom 
is not true \cite{prd80,prl81}. It tells that QCD has 
two types of gluons, the color neutral neurons and 
colored chromons,  which play totally different roles. 
Moreover, it has a non-trivial core, the restricted 
QCD (RCD) which describes the Abelian sub-dynamics 
of QCD but has the full non-Abelian gauge symmetry, 
which we can separate from QCD gauge independently. 

There are ample motivations for the Abelian decomposition. Consider the proton made of three quarks. Obviously 
we need the gluons to bind the quarks in the proton. 
However, the quark model tells that the proton has no 
valence gluon. If so, what is the binding gluon which 
bind the quarks in proton, and how do we distinguish 
it from the valence gluon? 

Another motivation is the color confinement in QCD. 
Two popular proposals to resolve this problem are 
the monopole condensation \cite{prd80,prl81,nambu} 
and the Abelian dominance \cite{thooft,prd00}.  To 
prove the monopole condensation, we first have to 
separate the monopole potential gauge independently. 
Similarly, to prove the Abelian dominance we have to 
know what is the Abelian part and how to separate it. 
How can we do that?

The Abelian decomposition decomposes the non-Abelian 
gauge potential to two parts, the restricted Abelian part 
which has the full non-Abelian gauge symmetry and 
the gauge covariant non-Abelian part which describes 
the colored valence gluons (the chromons) \cite{prd80,prl81}. Moreover, it separates the restricted potential to 
the non-topological Maxwell part which describes 
the colorless binding gluons (the neurons) and 
the topological Dirac part which describes 
the non-Abelian monopole \cite{prd80,prl81}. 

This has deep consequences. It tells that there 
is a simpler QCD called the restricted QCD (RCD) 
made of the restricted potential which describes 
the Abelian sub-dynamics of QCD, and that QCD 
can be viewed as RCD which has the chromons 
as the colored source. More importantly, it tells 
that QCD has two types of gluons which play totally 
different roles.  

This allows us to prove the Abelian dominance, that RCD 
is responsible for the confinement \cite{thooft,prd00}. 
Since the colored chromons have to be confined, it 
can not play any role in the confinement. Moreover, 
this allows us to prove the monopole condensation. 
Integrating out the chromons under the monopole 
background, we can demonstrate that the true QCD 
vacuum is given by stable monopole condensation \cite{prd13,ijmpa14,arx18}. This makes the experimental verification of the Abelian decomposition, the confirmation 
of two types of gluons, an urgent issue. The purpose of 
this letter is to discuss how to do this at LHC.
    
When the Abelian decomposition was proposed first, 
there was no way to verify this experimentally. During 
the last twenty years, however, there has been huge 
progress on jet physics. Theoretically new features 
of the jet substructure have been known which can 
tag the different jets \cite{jt1,jt2,jt3,jt4,jt5}. Moreover, 
ATLAS and CMS have succeeded to separate different 
types of jet experimentally \cite{je1,je2,je3}. We argue 
that these progresses could allow us to confirm 
the existence of two types of gluon jets experimentally 
at LHC.  

Before we do this, we have to know how QCD has 
two types of gluons. To show this consider the SU(2) 
QCD first, and select the Abelian direction $\n$. 
Impose the isometry to project out the  restricted 
potential $\hA_\mu$ \cite{prd80,prl81}
\begin{gather}
D_\mu \n=(\pd_\mu+g\A_\mu \times) \n=0,  \nn\\
\A_\mu \rightarrow \hA_\mu
=A_\mu \n-\frac1{g} \n \times \pd_\mu \n
=\tA_\mu+\tC_\mu,  \nn\\
\tA_\mu=A_\mu \n,~~A_\mu=\n \cdot \A_\mu,  
~~\tC_\mu=-\frac1{g} \n \times \pd_\mu \n.
\label{ap}
\end{gather}
It is made of two parts, the non-topological 
(Maxwellian) $\tA_\mu$ which describes the color 
neutral Abelian gluon (the neuron) and the topological 
(Diracian) $\tC_\mu$ which describes the non-Abelian 
monopole \cite{prl80,plb82}. With this we have
\begin{gather}
\hF_\mn = (F_\mn+ H_\mn) \n=F'_\mn \n, \nn\\
F_\mn = \pd_\mu A_\nu-\pd_\nu A_\mu,  \nn\\
H_\mn = -\frac1g  \n \cdot (\pd_\mu \n\times \pd_\nu \n)
=\pd_\mu C_\nu-\pd_\nu C_\mu,  \nn\\
C_\mu=-\frac1g \n_1 \cdot \pd_\mu \n_2,  \nn\\
F'_\mn=\pd_\mu A'_\nu-\pd_\nu A'_\mu,
~~~A'_\mu=A_\mu+C_\mu.
\label{rf} 
\end{gather}
Notice that $\hF_\mn$ is Abelian but is made 
of two potentials, the non-topological $A_\mu$ and 
topological $C_\mu$. With (\ref{ap}) we can construct 
RCD which has the full non-Abelian gauge symmetry,
\begin{gather}
{\cal L}_{RCD} =-\frac14 \hF^2_\mn=-\frac14 F_\mn^2 \nn\\
+\frac1{2g} F_\mn \n \cdot (\pd_\mu \n \times \pd_\nu \n)
-\frac1{4g^2} (\pd_\mu \n \times \pd_\nu \n)^2,
\label{rcd}
\end{gather}
which describes the Abelian sub-dynamics of QCD. 

We can express the full SU(2) potential adding 
the non-Abelian part $\X_\mu$ to 
$\hA_\mu$ \cite{prd80,prl81}
\bea
&\A_\mu = \hA_\mu + \X_\mu,    
~~~~\n \cdot \X_\mu=0,
\label{adec}
\eea
and show that $\hA_\mu$ has the full gauge degrees 
of freedom while $\X_\mu$ transforms gauge 
covariantly. Moreover, with
\bea
\F_\mn=\hF_\mn + \hD _\mu \X_\nu 
- \hD_\nu \X_\mu + g\X_\mu \times \X_\nu,
\eea 
we obtain the extended QCD (ECD) to recover the full 
QCD
\bea 
&{\cal L}_{ECD} = -\dfrac{1}{4} \F^2_\mn
=-\dfrac{1}{4} \hF_\mn^2-\dfrac{1}{4}(\hD_\mu\X_\nu
-\hD_\nu\X_\mu)^2 \nn\\
&-\dfrac{g}{2} {\hat F}_\mn \cdot (\X_\mu \times \X_\nu)
-\dfrac{g^2}{4} (\X_\mu \times \X_\nu)^2, 
\label{2ecd} 
\eea
This shows that QCD can be viewed as RCD which has 
the chromon as the colored source \cite{prd80,prl81}. 

The Abelian decomposition of SU(3) QCD is 
similar \cite{prd13,ijmpa14,arx18}. Let $\n_i~(i=1,2,...,8)$ 
be the orthonormal SU(3) basis and choose $\n_3=\n$ 
and $\n_8=\n'$ to be the Abelian directions,
and impose the isometry
\begin{gather}
D_\mu \n=0,~~~~D_\mu \n'=0.
\label {3cp}
\end{gather}
Solving this we have the SU(3) Abelian 
projection,
\begin{gather}
\A_\mu \rightarrow \hA_\mu=A_\mu \n+A_\mu' \n'
-\frac1g \n\times \pd_\mu \n-\frac1g \n'\times \pd_\mu \n' \nn\\
=\sum_p \frac23 \hA_\mu^p,~~~(p=1,2,3),    \nn\\
\hA_\mu^p=A_\mu^p \n^p-\frac1g \n^p 
\times \pd_\mu \n^p=\cA_\mu^p+\cC_\mu^p, \nn\\
A_\mu^1=A_\mu,
~~~A_\mu^2=-\frac12 A_\mu
+\frac{\sqrt 3}{2} A_\mu',  \nn\\
A_\mu^3=-\frac12 A_\mu-\frac{\sqrt 3}{2}A_\mu', 
~~~\n^1=\n,   \nn\\
\n^2=-\frac12 \n +\frac{\sqrt 3}{2} \n', 
~~~\n^3=-\frac12 \n -\frac{\sqrt 3}{2} \n',
\label{cp3}
\end{gather}
where $A_\mu$ and $A_\mu'$ are two SU(3) neurons, 
and the sum is the sum of the three Abelian directions 
$(\n^1,\n^2,\n^3)$ of three SU(2) subgroups. Notice 
that $\hA_\mu$ is expressed by three SU(2) restricted 
potential $\hA_\mu^i~(i=1,2,3)$ in Weyl symmetric 
way. From this we have the SU(3) RCD
\begin{gather}
{\cal L}_{RCD} = -\frac14 \hF_\mn^2
=-\sum_p \frac16 (\hF_\mn^p)^2, 
\label{rcd3}
\end{gather}
which has the full SU(3) gauge symmetry. 

Adding the valence part $\X_\mu$ to $\hA_\mu$ we have 
the SU(3) Abelian decomposition,
\begin{gather}
\A_\mu=\hA_\mu+\X_\mu
=\sum_p (\dfrac23 \hA_\mu^p+\W_\mu^p), \nn\\
\W_\mu^1= X_\mu^1 \n_1+ X_\mu^2 \n_2,
~~~\W_\mu^2=X_\mu^6 \n_6 + X_\mu^7 \n_7,  \nn\\
\W_\mu^3= X_\mu^4 \n_4  +X_\mu^5 \n_5.
\label{cdec3}
\end{gather}
Notice that $\X_\mu$ can be decomposed to three 
(red, blue, and green) SU(2) chromons 
$(\W_\mu^1,\W_\mu^2,\W_\mu^3)$. 

\begin{figure}
\includegraphics[scale=0.65]{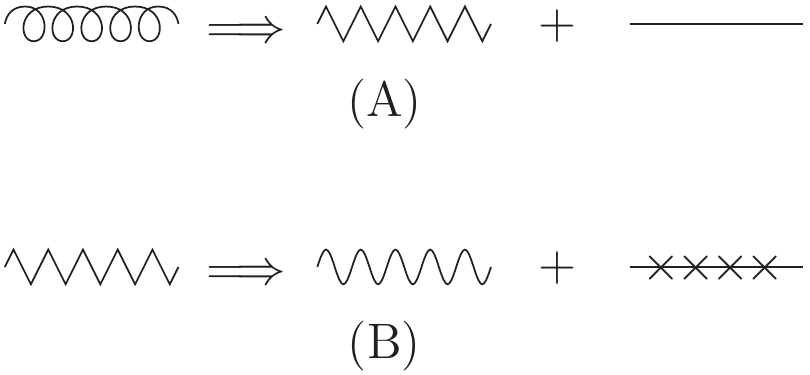}
\caption{\label{cdec} The Abelian decomposition 
of the gauge potential. In (A) it is decomposed to 
the restricted potential (kinked line) and the chromon
(straight line). In (B) the restricted potential is 
further decomposed to the neuron (wiggly line) and 
the monopole (spiked line).}
\end{figure}

From this we have
\begin{gather}
\F_\mn=\hF_\mn + \hD _\mu \X_\nu 
-\hD_\nu \X_\mu + g\X_\mu \times \X_\nu  \nn\\
=\sum_p \big[\frac23 \hF_\mn^p
+ (\hD_\mu^p \W_\nu^p-\hD_\mu^p \W_\nu^p) \big] 
+\sum_{p,q}\W_\mu^p \times \W_\nu^q,   \nn\\
\hD_\mu^p=\pro_\mu+ g \hA_\mu^p \times,
\end{gather}
and obtain the SU(3) ECD \cite{prd13,ijmpa14,arx18}
\begin{gather}
{\cal L}_{ECD}= -\frac14 \F_\mn^2
=\sum_p \Big\{-\dfrac{1}{6} (\hF_\mn^p)^2 \nn\\
-\frac14 (\hD_\mu^p \W_\nu^p- \hD_\nu^p \W_\mu^p)^2 
-\frac{g}{2} \hF_\mn^p \cdot (\W_\mu^p 
\times \W_\nu^p) \Big\} \nn\\
-\sum_{p,q} \frac{g^2}{4} (\W_\mu^p 
\times \W_\mu^q)^2 \nn\\
-\sum_{p,q,r} \frac{g}2 (\hD_\mu^p \W_\nu^p
- \hD_\nu^p \W_\mu^p) \cdot (\W_\mu^q 
\times \W_\mu^r)  \nn\\
-\sum_{p\ne q} \dfrac{g^2}{4} \Big[(\W_\mu^p 
\times \W_\nu^q)
\cdot (\W_\mu^q \times \W_\nu^p)  \nn\\
+(\W_\mu^p \times \W_\nu^p)\cdot (\W_\mu^q 
\times \W_\nu^q) \Big].
\label{3ecd}
\end{gather}
The Abelian decomposition is known as the Cho 
decomposition, Cho-Duan-Ge (CDG) decomposition, 
or Cho-Faddeev-Niemi (CFN) decomposition \cite{fadd,shab,zucc,kondo,kondor}.

We can add quarks in the Abelian decomposition,
\begin{gather}
{\cal L}_{q} =\bPsi (i\gamma^\mu D_\mu-m) \Psi \nn\\
= \bPsi (i\gamma^\mu \hD_\mu-m) \Psi 
+\frac{g}{2} \X_\mu \cdot \bPsi (\gamma^\mu \vlam) \Psi  \nn\\
=\sum_p \Big[\bPsi^p (i\gamma^\mu \hD_\mu^p-m) \Psi^p
+\frac{g}{2} \W_\mu^p \cdot \bPsi^p
(\gamma^\mu \vsig) \Psi^p \Big], \nn\\
\hD_\mu = \pd_\mu +\frac{g}{2i} {\vlam}\cdot \hA_\mu,  
~~~\hD_\mu^p=\pd_\mu
+\frac{g}{2i} {\vsig}\cdot \hA_\mu^p,
\label{qlag}
\end{gather}
where $m$ is the mass, $p$ denotes the color of 
the quarks, and $\Psi^p$ represents the three SU(2) 
quark doublets $(r,b)$, $(b,g)$, and $(g,r)$. Notice 
that the Lagrangian is also Weyl-symmetric.

There has been an assertion that the introduction 
of the Abelian direction $\n$ adds a new dynamical 
degree \cite{fadd}. This is a gross misunderstanding. 
It represents the gauge degree, not a dynamical 
degree \cite{prd01}. Nevertheless $\n$ plays important 
roles representing the topological structure of QCD, 
the monopole topology $\pi_2(S^2)$ as well as 
the vacuum topology $\pi_3(S^3)\simeq\pi_3(S^2)$, 
both of which are the essential characteristics of 
QCD \cite{prl80,plb82}. 

\begin{figure}
\includegraphics[height=4cm, width=7cm]{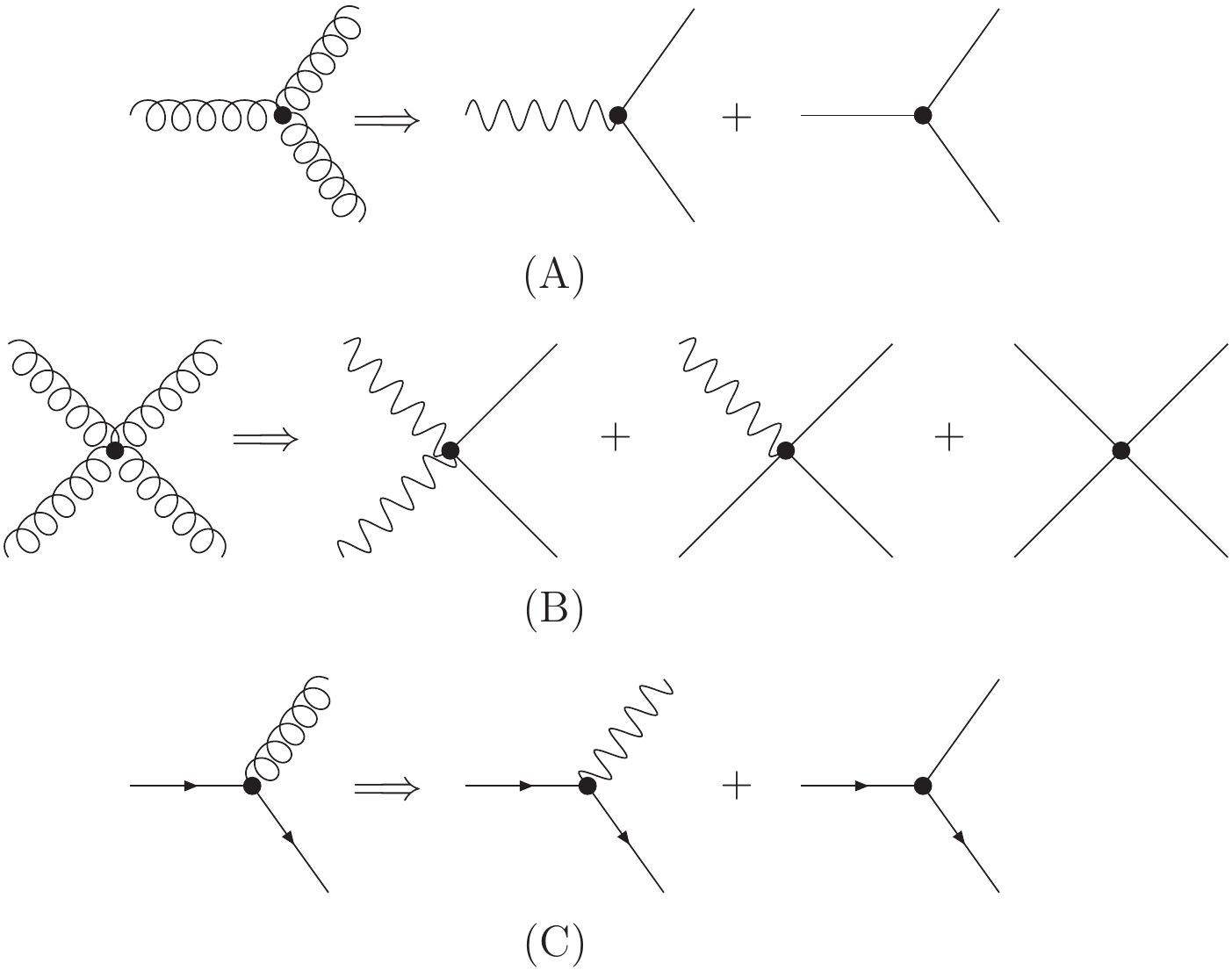}
\caption{\label{3ecdint} The decomposition of 
the Feynman diagrams in SU(3) QCD. In (A) and (B)
the three-point and four-point gluon vertices are 
decomposed, and in (C) the quark-gluon vertices 
are decomposed.}
\end{figure}

The Abelian decomposition is expressed graphically 
in Fig. \ref{cdec}. Although the decomposition does 
not change QCD, it reveals the important hidden 
structures of QCD. In particular, it shows the existence 
of two types of gluons, the neuron and chromon, which 
play totally different roles. This is evident in (\ref{2ecd}), (\ref{3ecd}), and (\ref{qlag}). The neurons, just like 
the photons in QED, provide the binding. But the chromons, 
just like the quarks, become the colored source. 

This has deep implications. In the perturbative regime 
this tells that the Feynman diagram can be decomposed 
in such a way that the color conservation is explicit. 
This is graphically shown in Fig. \ref{3ecdint}. Notice 
that the monopole does not appear in the Feynman 
diagram, because it does not represent a dynamical 
degree. 

\begin{figure}
\includegraphics[height=4.5cm, width=6cm]{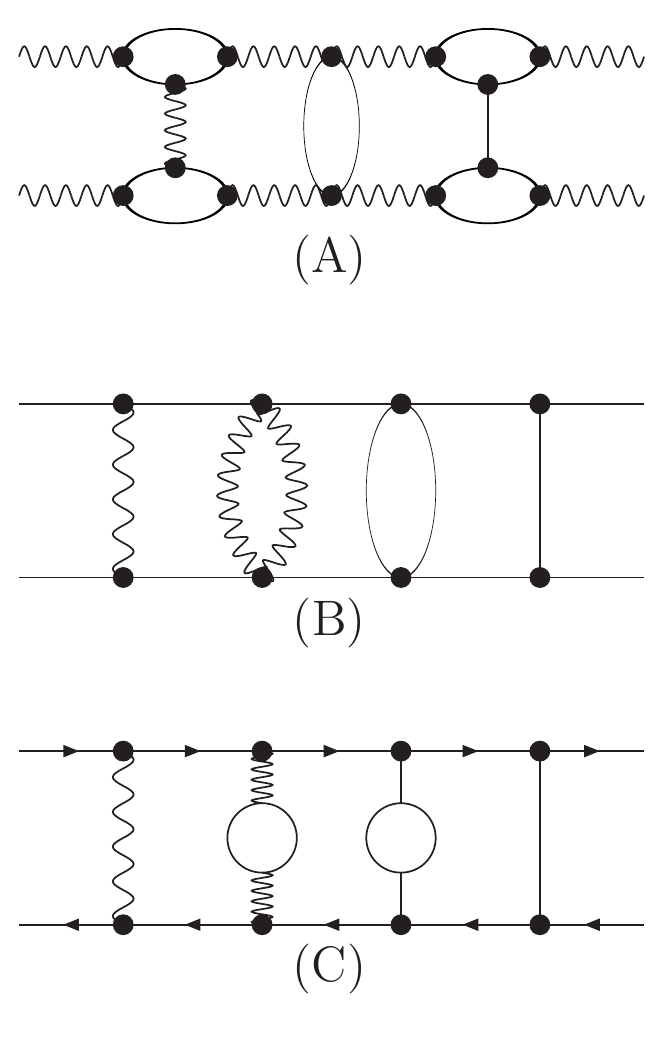}
\caption{\label{bind} The possible Feynman diagrams 
of the neurons and chromons. Two neuron interaction 
is shown in (A), two chromon interaction is shown in (B), 
and quark-antiquark interaction is shown in (C).}
\end{figure}

To see the decomposition is non-trivial, consider 
the Feynman diagrams of two neurons, two chromons, 
and quark-antiquark pair shown in Fig. \ref{bind}. 
Clearly the neuron binding looks very much like 
two photon binding in QED, while the chromon binding 
look just like the quark-antiquark binding in QCD. 
This is because the three point vertex made of two 
or three neurons are forbidden. This strongly implies 
that the neurons can hardly make a bound state, 
and may not be viewed as the constituent of hadrons. 
However, the chromon binding strongly implies that 
they, just like the quarks, become the constituent 
of hadrons and form hadronic bound states. This 
generalizes the quark model to the quark and 
chromon model, which provides a new picture of 
hadrons \cite{prd15,prd18}. 

In particular, this gives us a new picture of glueball 
different from all existing glueball models. In existing 
glueball models all gluons are treated equally, and 
the color singlet combinations of the gluon octet 
make the glueballs. But in the quark and chromon 
model only the chromons become the constituent 
of glueballs, and the neurons provide the chromon 
binding \cite{prd80,prl81}. Moreover, this picture 
describes the glueball-quarkonium mixing successfully. 
The numerical analysis of the mixing in this picture 
below 2 GeV shows that $f_0(1500)$, $f_2(1950)$,
$\eta(1405)$, and $\eta(1475)$ in $0^{++}$, $2^{++}$, 
and $0^{-+}$ sectors can be identified as predominantly 
the glueball states \cite{prd15,prd18}. 

In the non-perturbative regime, the Abelian decomposition 
proves the Abelian dominance. Implementing the Abelian decomposition on lattice we can calculate the Wilson 
loop integral with the full potential, the restricted potential, 
and the monopole potential separately, and show that 
all three potentials produce exactly the same linear 
confining force \cite{kondo,kondor,cundy}. This proves 
not only the Abelian dominance but also the monopole dominance. 

As importantly the Abelian decomposition puts QCD 
to the background field formalism, because we can 
treat the restricted part as the classical background 
and the valence part as the fast moving quantum 
field \cite{dewitt,prd01}.  This makes ECD an ideal 
platform for us to calculate the QCD effective 
potential and prove the monopole condensation. 
Indeed, choosing the monopole potential as 
the background and integrating out the chromons 
gauge invariantly, we can prove that the true QCD 
vacuum is given by the gauge invariant monopole 
condensation \cite{prd13,ijmpa14,arx18}.   

The prediction and subsequent confirmation of the gluon 
jet was a great success of QCD \cite{glujet,eje1,eje2}. 
To promote QCD further, what we badly need is 
the experimental confirmation of the Abelian 
decomposition (the existence of two types of gluons). 
This could be done at LHC. Obviously two types of 
gluons mean two types of gluon jets, the neuron jet 
and chromon jet, and the recent progress on jet 
physics could allow us to separate the neuron jet from 
the chromon jet. This could be a difficult task, but 
ATLAS and CMS have already separated the gluon 
jets from the quark jets successfully \cite{je1,je2,je3}. 
In doing so they have developed powerful techniques 
to tag different jets. This could allow them to identify 
the neuron jet.  

\begin{figure}
\psfig{figure=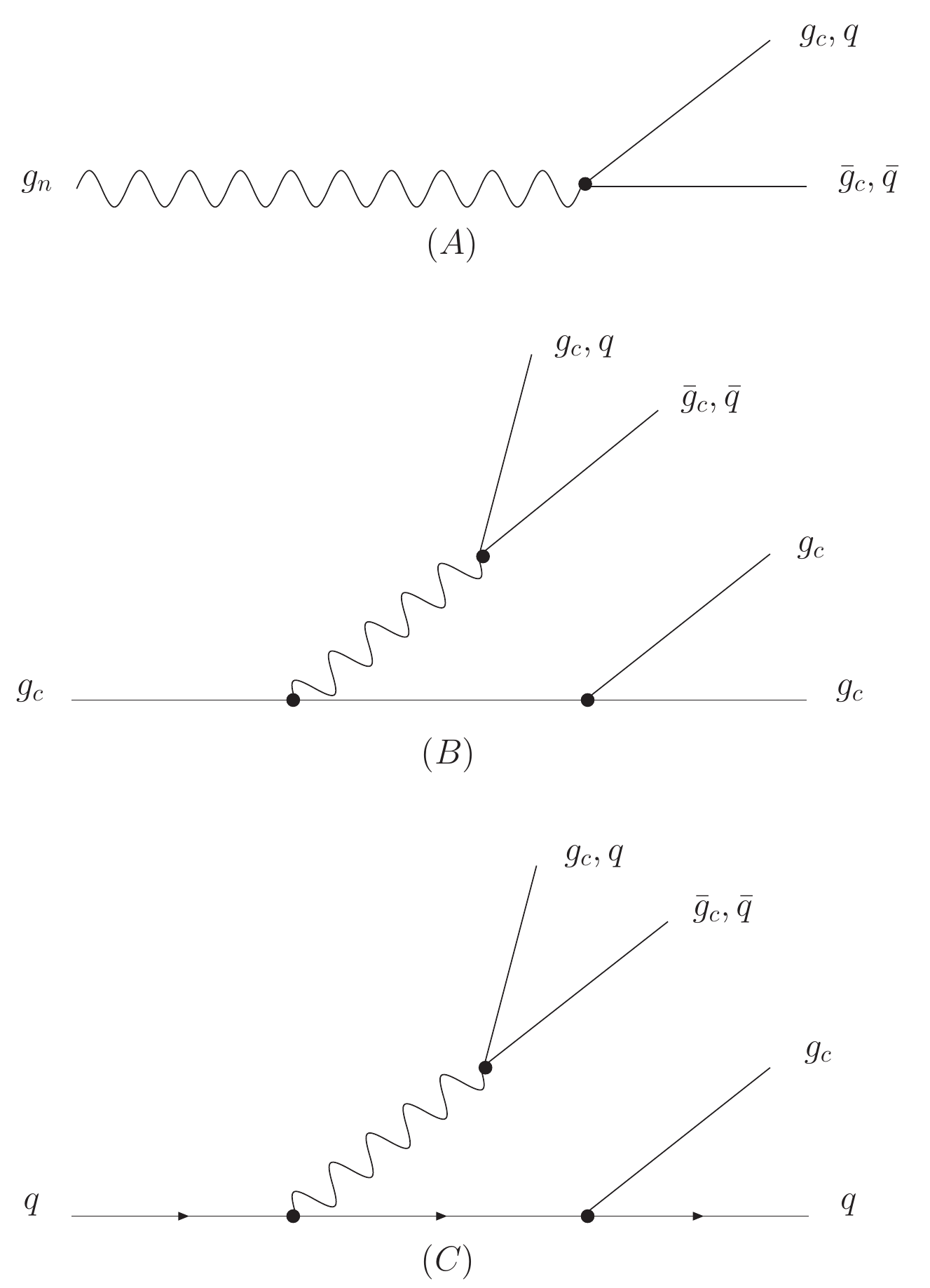, height=4.5cm, width=7.5cm}
\caption{\label{3jet} The perturbative diagrams of neuron, chromon, and quark jets. The neuron jet described in 
(A) is qualitatively different from the chromon jet and 
the quark jet shown in (B) and (C), while the chromon 
and quark jets are similar.}
\end{figure}

To show this might be possible, we have to know what 
are the expected differences of the neuron jet from 
other jets. The gluons and quarks emitted in the p-p 
collisions evolve into hadron jets in two steps, the parton 
shower described by the perturbative process and 
the hadronization described by the non-perturbative 
process. The neuron behaves differently in the first step. 
This is shown in Fig. \ref{3jet}. Clearly the neuron jet 
shown in (A) has no parton shower made of neuron 
emission which exists in the chromon jet (B) and 
the quark jet (C), but the chromon and quark jets 
look almost the same. This is because the three-point 
vertex made of three neurons is forbidden. This strongly 
suggests that the neuron jet must have different jet 
shape, sharp with relatively small radius compared to 
the chromon and quark jets. 

There are other differences. The neuron jet should have 
different (charged) particle multiplicity, considerably 
smaller than that of the quark and/or chromon jets. 
This again must be clear from Fig. \ref{3jet}, which shows 
that chromon and quark couple to neurons which 
make secondary showers but the neuron jet is not 
allowed to have such interaction.  

To quantify the differences, we have to know the color 
factor of the neuron and chromon jets, a most important 
quantity which determines the characters of the jet.
Although the neurons are color neutral, they have a finite 
color factor. But we can not find it from the SU(3) 
Casimir invariant, because the color gauge symmetry 
is replaced to the 24-element color reflection symmetry 
after the Abelian decomposition \cite{prd80,prl81}. 
There are various ways to find the neuron color factor, 
but from the fact that the adjoint representation has 
the color factor 3, we can deduce the neuron color 
factor to be 3/4. This comes from the simple democracy 
of the gauge interaction. Since the neurons constitute 
one quarter of total gluons their color factor becomes 
3/4 and that of chromons becomes 9/4, while that of 
of quarks remains the same. This endorses the above 
predictions.

Another important feature of the neuron jet is that it
has has different color flow. Clearly the chromons and 
quarks carry color charge, but the neurons are color 
neutral. So the neuron jet must have different color 
flow. In fact, Fig. \ref{3jet} tells that the color flow of 
the neuron jet generates an ideal color dipole pattern, 
but the other two jets have distorted dipole pattern. We 
could check this prediction using Pythia and FastJet. 
This means that the neuron jet must be quantitatively 
different from the chromon and quark jets. To tell more 
detailed differences, of course, we need more serious 
theoretical calculations. But the above differences 
give us enough tools to identify the neuron jet.  

At this point one might ask what are the gluon jets 
identified by ATLAS and CMS? Probably they are 
the chromon jets, because the chromon jet has 
the characteristics of the known gluon jet. This is 
evident from Fig. \ref{3jet}. Perhaps a more interesting 
question is why they have not found the neuron jet? 
There could be two explanations. They have not 
searched for the neuron jet yet, because they had 
no motivation to do that. Or they might have misidentified 
some of the neuron jets as the quark jet. This is because 
the color factor of neuron and chromon jets are not 
much different. This tells that we need a more careful
analysis of quark and gluon jets.   

If this is true, the recent experiments which separated 
the quark jet from the gluon jet based on the color 
factor ratio $C_A/C_F=9/4$ need to be completely 
re-analysed \cite{je1,je2,je3}. According to the above 
reasoning we should look for three (neuron, quark, 
and chromon) jets which have the color factor ratio 
$3/4:4/3:9/4\simeq 0.56:1:1.69$. In this respect we 
notice two interesting reports which might indicate 
that the observed gluon jets are indeed the chromon 
jets. A re-analysis of DELPHI $e^+e^-$ three jet data 
at LEP strongly suggests that actual $C_A/C_F$ could 
be around 1.74, much less than the popular value 2.25 
but close to our prediction 1.69 \cite{lep,eje3}. 
Moreover, the $p \bp$ D{\O} jets experiment at Fermilab 
Tevatron shows that the quark to gluon jets particle 
multiplicity ratio is around 1.84, again close to our 
prediction \cite{do}. They seem to imply that the above 
interpretation might be correct.  

One advantage in searching for the neuron jet is that 
we do not need any new collider or detector. LHC 
produces billions of hadron jets in a second, and ATLAS 
and CMS have already filed up huge data on jets. So 
all that we have to do is to re-analyse these data. 
Actually we could even go back to the three jet events 
(the gluon jets) at DESY, LEP, and Tevatron, and 
search for the neuron jets \cite{eje1,eje2,eje3,eje4,do}. 
Here again the simple number counting strongly 
suggests that one-quarter of the gluon jets coming 
from the three jets events could actually be the neuron 
jets. It would be very interesting to re-analyse 
the existing data and confirm the existence of the neuron 
jet.  

The confirmation of the gluon jet justified the asymptotic 
freedom and extended our understanding of QCD very 
much \cite{wil}. Clearly the experimental confirmation 
of two types of gluon jets would be at least as important. 
It will shed a new light on QCD, revealing the hidden 
structures of QCD. In particular it will endorse 
the decomposition of Feynman diagram and justify 
the quark and chromon model \cite{prd15,prd18}. 
The detailed discussion of the neuron jet and its color 
flow will be published separately \cite{cho}. 

{\bf Acknowledgements}

~~~The work is supported in part by National Natural 
Science Foundation of China (Grant 11575254), National 
Research Foundation of Korea funded by the Ministry 
of Education (Grants 2015-R1D1A1A0-1057578 and 
2018-R1D1A1B0-7045163), and by the Center for 
Quantum Spacetime at Sogang University.


\begin{thebibliography}{99}
\bibitem{prd80} Y. M. Cho, Phys. Rev. {\bf D21}, 1080 (1980);
Y. S. Duan and M. L. Ge, Sci. Sinica {\bf 11}, 1072 (1979).
\bibitem{prl81} Y. M. Cho, Phys. Rev. Lett. {\bf 46}, 
302 (1981); Phys. Rev. {\bf D23}, 2415 (1981).
\bibitem{nambu} Y. Nambu, Phys. Rev. {\bf D10}, 4262 
(1974); S. Mandelstam, Phys. Rep. {\bf 23C}, 245 (1976);
A. Polyakov, Nucl. Phys. {\bf B120}, 429 (1977).
\bibitem{thooft} G. 't Hooft, Nucl. Phys. {\bf B190}, 
455 (1981).
\bibitem{prd00} Y. M. Cho, Phys. Rev. {\bf D62}, 
074009 (2000). 
\bibitem{prd13} Y. M. Cho, Franklin H. Cho, and J. H. Yoon, 
Phys. Rev. {\bf D87}, 085025 (2013).
\bibitem{ijmpa14} Y. M. Cho, Int. J. Mod. Phys. {\bf A29}, 
1450013 (2014); Y. M. Cho, Euro Phys. J. WoC, {\bf 182}, 
02029 (2018).
\bibitem{arx18} Y. M. Cho, and Franklin H. Cho, 
submitted to Phys. Rev. {\bf D}.

\bibitem{jt1} H. Nilles and K. streng, Phys. Rev. {\bf D23},
1944 (1981);L. Jones, Phys. Rev. {\bf D39}, 2550 (1989).
\bibitem{jt2} Z. Fodor, Phys. Rev. {\bf D41},1726 (1990);
L. jones, Phys. Rev. {\bf D42},811 (1990); L. Lonnblad, 
C. Peterson, and T. Rognvaldsson, Nucl. Phys. {\bf B349}, 
675 (1991); J. Pumplin, Phys. Rev. {\bf D44}, 2025 (1991).
\bibitem{jt3} J. Gallicchio and M. Schwartz, Phys. Rev. 
Lett. {\bf 107}, 172001 (2011); A. Larkoski, G. Salam, and
J. Thaler, JHEP, {\bf 06}, 108 (2013); 
\bibitem{jt4} B. Bhattacherjee, S. Mukhopadhyay, M. Nojiri, 
Y. Sakaki, and B. Webber, JHEP, {\bf 04}, 131 (2015); 
D. de Lima, P. Petrov, D. Soper, and m. Spannowsky, 
Phys. Rev. {\bf D95}, 034001 (2017); 
\bibitem{jt5} J. Davighi and P. Harris, Eur. Phys. J. {\bf C78}, 
334 (2018); E. Metodiev and J. Taler, Phys. Rev. Lett. {\bf120}, 
241602 (2018).

\bibitem{je1} ATLAS Collaboration, Eur. Phys. J. {\bf C73}, 
2676 (2013); {\bf C74}, 3023 (2014); {\bf C75}, 17 (2015). 
\bibitem{je2} CMS Collaboration, Eur. Phys. J. {\bf C75}, 
66 (2015); Phys. Rev. {\bf D92}, 032008 (2015). 
\bibitem{je3} ATLAS Collaboration, Eur. Phys. J. {\bf C76}, 
322 (2016); Phys. Rev. {\bf d96}, 072002 (2017). 

\bibitem{prl80} Y. M. Cho, Phys. Rev. Lett. {\bf 44}, 1115 (1980).
\bibitem{plb82} Y. M. Cho, Phys. Lett. {\bf B115}, 125 (1982).

\bibitem{fadd} L. Faddeev and A. Niemi, Phys. Rev. Lett.
{\bf 82}, 1624 (1999); Phys. Lett. {\bf B449}, 214 (1999).
\bibitem{shab} S. Shabanov, Phys. Lett. {\bf B458}, 322 
(1999); {\bf B463}, 263 (1999); H. Gies, Phys. Rev. {\bf D63}, 
125023 (2001).
\bibitem{zucc} R. Zucchini, Int. J. Geom. Meth. Mod. Phys. 
{\bf 1}, 813 (2004).
\bibitem{kondo} S. Kato, K. Kondo, T. Murakami, A. Shibata, 
T. Shinohara, and S. Ito, Phys. Lett. {\bf B632}, 326 (2006); 
{\bf B645}, 67 (2007); {\bf B653}, 101 (2007); {\bf B669}, 
107 (2008). 
\bibitem{kondor} K. Kondo, S. Kato, A. Shibata, 
and T. Shinohara, Phys. Rep. {\bf 579}, 1 (2015).

\bibitem{prd15} Y. M. Cho, X. Y. Pham, Pengming Zhang, 
Ju-Jun Xie, and Li-Ping Zou, Phys. Rev. {\bf D91}, 114020 
(2015); Y. M. Cho, Euro Phys. J. WoC, {\bf 182}, 02031 (2018).
\bibitem{prd18} Pengming Zhang, Li-Ping Zou, and Y. M. Cho, 
Phys. Rev. {\bf D98}, 096015 (2018). 

\bibitem{cundy} N. Cundy, Y. M. Cho, W. Lee, 
and J. Leem, Phys. Lett. {\bf B729}, 192 (2014); 
Nucl. Phys. {\bf B895}, 64 (2015).

\bibitem{dewitt} B. de Witt, Phys. Rev. {\bf 162}, 1195 (1967);
1239 (1967).
\bibitem{prd01} W. S. Bae, Y. M. Cho, and S. W. Kim, 
Phys. Rev. {\bf D65}, 025005 (2001).

\bibitem{glujet} J. Ellis, M.K. Gaillard, and G.G. Ross,
Nucl. Phys. {\bf B111}, 253 (1976).
\bibitem{eje1} R. Brandelik et al. (TASSO Collaboration),
Phys. Lett. {\bf B86}, 243 (1979); D.P. Barber et al. 
(MARK-J Collaboration), Phys. Rev. Lett. {\bf 43}, 830 (1979).
\bibitem{eje2} Ch. Berger et al. (PLUTO Collaboration), 
Phys. Lett. {\bf B86}, 418 (1979); W. Bartel et al. 
(JADE Collaboration), Phys. Lett. {\bf B91}, 142 (1980).
\bibitem{lep} J. Gary, Phys. Rev. {\bf D61}, 114007 (2000).
\bibitem{eje3} P. Abreu et al. (DELPHI Collaboration), 
Phys. Lett. {\bf B449}, 383 (1999).
\bibitem{do} V. Avazov et al. (D{\O} Collaboration), Phys. 
Rev. {\bf D65}, 052008 (2002).
\bibitem{eje4} G. Abbiendi et al. (OPAL Collaboration),
Euro.Phys. J. {\bf C11}, 217 (1999).

\bibitem{wil} D. Gross and F. Wilczek, Phys. Rev. Lett. {\bf 30},
1343 (1973); H. Politzer, Phys. Rev. Lett. {\bf 30}, 1346 (1973).

\bibitem{cho} Y. M. Cho, Xiaohui Liu, and Pengming Zhang, 
to be published.

\end{thebibliography}
\end{document}